\documentclass[%
reprint,
superscriptaddress,
nofootinbib,
longbibliography,
amsmath,
amssymb,
aps,
showkeys
]{revtex4-2}
\usepackage{graphicx}
\usepackage{dcolumn}
\usepackage{bm}
\usepackage{hyperref}
\usepackage[bottom]{footmisc}
\usepackage{footnote}
\usepackage{xcolor}
\usepackage[mathlines]{lineno}
\usepackage{fontenc}
\usepackage{graphicx}
\usepackage{dcolumn}
\usepackage{bm}
\usepackage{booktabs}
\usepackage{array,multirow}
\usepackage{amsmath}
\usepackage{amsfonts}
\usepackage{amssymb}
\usepackage{mathrsfs}
\usepackage{enumerate}
\usepackage{fancyhdr}
\usepackage{xcolor}
\usepackage{graphicx}
\usepackage{listings}
\usepackage{float}
\usepackage{verbatim}
\usepackage{braket}
\usepackage{centernot}
\usepackage{setspace}
\usepackage{endnotes}
\usepackage[normalem]{ulem}
\usepackage{wrapfig}
\usepackage{multirow}
\usepackage{comment}
\usepackage{kantlipsum}
\allowdisplaybreaks
\usepackage{lipsum, babel}
\hypersetup{breaklinks = true, colorlinks = true, citecolor = blue, linkcolor = blue, urlcolor = blue}
\usepackage[mathlines]{lineno}

\begin{document}

\title{Global $\Lambda$ hyperon polarization in low-energy heavy ion collisions \\- a scenario without vorticity}

\author{Feng Liu}
\email{feng.liu.2@stonybrook.edu}
\affiliation{Department of Physics and Astronomy, Stony Brook University, New York 11794-3800, USA}
\author{Zhoudunming~Tu}
\email{zhoudunming@bnl.gov}
\affiliation{Department of Physics and Astronomy, Stony Brook University, New York 11794-3800, USA}
\affiliation{Department of Physics, Brookhaven National Laboratory, Upton, New York 11973-5000, USA}

\date{\today}
             
\begin{abstract} 
Since its discovery, global polarization of the $\Lambda$ hyperon in heavy-ion collisions has been firmly established and is widely attributed to the large vorticity generated in the rotating quark–gluon plasma. In contrast, nearly fifty years after the first observation of unexpectedly large transverse $\Lambda$ polarization in unpolarized hadron collisions, its underlying mechanism remains an open and long-standing puzzle, despite being observed across a broad range of collision systems. Although these two phenomena exhibit notable similarities, they are generally regarded as arising from distinct physical origins. In this work, we propose a direct connection between $\Lambda$ global polarization in heavy-ion collisions and the long-standing transverse polarization observed in unpolarized collision systems. We demonstrate that the alignment between the $\Lambda$ production plane and the reaction plane, driven by directed flow, can transfer transverse polarization into the measured global polarization signal. Realistic Monte Carlo simulations of Au+Au collisions at $\sqrt{s_{\rm NN}} = 3~\text{GeV}$ indicate that this mechanism can generate a sizable global polarization, accounting for approximately $23\%\pm6\%$ of the magnitude reported by the STAR Collaboration. Our results establish, for the first time, a quantitative link between these two well-known phenomena and have important implications for the interpretation of $\Lambda$ global polarization measurements in low-energy heavy-ion collisions.
\end{abstract}

\maketitle
\section{Introduction}

With the advent of the Relativistic Heavy Ion Collider (RHIC) and the Large Hadron Collider(LHC), high-energy collisions became available at unprecedented energies, providing ideal places for studies of spin phenomena~\cite{Nature2017,Nature2023,Nature2024,Nature2026,Abelev_2007,PhysRevC.98.014910,PhysRevC.104.L061901,STAR:2019erd,PhysRevC.108.014910,STAR:2025jwc,PhysRevD.91.032004,PhysRevC.101.044611,ALICE:2021pzu,CMS:2025nqr,PhysRevLett.123.132301,PhysRevLett.131.202301,PhysRevLett.126.162301,PhysRevLett.97.252001,PhysRevLett.100.232003,PhysRevD.86.032006,PhysRevLett.115.092002,PhysRevD.80.111108,PhysRevD.89.012001,PhysRevD.79.012003,PhysRevLett.103.012003,PhysRevD.90.012007,r5jw-lwhg,STAR:2025jwc,PhysRevD.100.072002,CMSCollaboration_2024,PhysRevD.110.112016}. Among them, the $\Lambda$ global polarization in heavy ion collisions (HICs) is a collective alignment of spins with respect to the direction of system's orbital angular momentum, which is perpendicular to the reaction plane. The enormous orbital angular momenta (OAM) stored in non-central collision systems are believed to be transferred, via spin-orbit coupling in the strongly rotating quark-gluon plasma (QGP), to spins of produced particles~\cite{Nature2017,PhysRevLett.94.102301}. Measurements~\cite{PhysRevC.101.044611,Abelev_2007,PhysRevC.98.014910,Nature2017,PhysRevC.98.014910,PhysRevC.104.L061901,PhysRevC.108.014910,Hu:2024suz} over the last two decades have confirmed a significant evolutionary trend of polarization magnitude with collision energy, i.e., the magnitude decreases as the collision energy increases.

Although studies of $\Lambda$ global polarization have been extremely successful, a conceptually similar but still unresolved phenomenon was already observed as early as 1976~\cite{PhysRevLett.36.1113}, known as the transverse $\Lambda$ polarization puzzle. Specifically, the $\Lambda$ hyperons produced from unpolarized proton-beryllium collisions were found to exhibit significant polarization with respect to their production planes. Such an observation is striking, as it stands in direct contradiction to the predictions of perturbative Quantum Chromodynamics (QCD), where large polarization cannot be naturally produced~\cite{PhysRevD.53.1073}. 
 Since then, transverse $\Lambda$  polarization has been extensively and systematically studied \cite{PhysRevD.9.608,PhysRevLett.36.1113,PhysRevLett.41.607,Amsterdam-CERN-Nijmegen-Oxford:1977qto,ERHAN1979301,Sugahara:1979wy,PhysRevLett.51.2025,AACHEN-BERLIN-BOMBAY-CERN-CRACOW-INNSBRUCK-JAMMU-LONDON-VIENNA-WARSAW:1985opf,PhysRevLett.56.2244,PhysRevD.40.3557,OPAL:1997oem,PhysRevD.43.2792,RAMBERG1994403,WA89:1994nzv,Fanti:1998px,ASTIER20003,ASTIER20013,WA89:2003ysd,ABT2006415,PhysRevD.76.092008,EPJC2007,SELEX:2007bjj,PhysRevD.90.072007,PhysRevD.91.032004,PhysRevLett.122.042001,LHCb:2024vwi,LHCb:2025rxf,BARNES1993277,PhysRevC.54.1877,PhysRevC.54.2831,PhysRevC.74.015206}, leading to well-established observations across many different collision systems, e.g., hadron-hadron and hadron-nucleus collisions \cite{PhysRevLett.36.1113,PhysRevLett.41.607,Amsterdam-CERN-Nijmegen-Oxford:1977qto,ERHAN1979301,PhysRevLett.51.2025,AACHEN-BERLIN-BOMBAY-CERN-CRACOW-INNSBRUCK-JAMMU-LONDON-VIENNA-WARSAW:1985opf,PhysRevLett.56.2244,PhysRevD.43.2792,RAMBERG1994403,WA89:2003ysd,ABT2006415,PhysRevD.40.3557,SELEX:2007bjj,STAR:2025jwc,BARNES1993277,PhysRevC.54.1877,PhysRevC.54.2831,PhysRevC.74.015206}, electron-nucleus collisions \cite{PhysRevD.76.092008,PhysRevD.90.072007}, electron-positron collisions \cite{PhysRevLett.122.042001}, and deep inelastic scatterings (DIS) \cite{ASTIER20003}. Although many theoretical efforts have been made to resolve this puzzle~\cite{PhysRevLett.68.907,PhysRevD.23.1227,PhysRevD.65.114014,PhysRevD.24.2419,Joseph:1981zv}, there is still no conclusive mechanism that can explain the global data at a quantitative level. 

Interestingly, theoretical efforts to understand the origin of transverse $\Lambda$  polarization inspired the early idea of $\Lambda$ global polarization in HICs~\cite{PhysRevLett.94.102301}, while the latter is widely believed to originate from partonic interactions during the evolution of the QGP, and is therefore, considered unrelated to the mechanism responsible for the transverse $\Lambda$ polarization. However, this picture is complicated by the fact that many existing explanations of the transverse $\Lambda$ polarization - such as parton recombination model~\cite{PhysRevD.23.1227,PhysRevD.24.2419}, interference from resonance decay~\cite{Joseph:1981zv}, single pion exchange~\cite{PhysRevLett.68.907}, and polarizing fragmentation function~\cite{PhysRevD.65.114014} - seem to point to a common origin: the nonperturbative hadronization or final state effects~\cite{PhysRevLett.68.907,PhysRevD.23.1227,PhysRevD.65.114014,PhysRevD.24.2419,Joseph:1981zv}. Since such effects are expected to occur universally, irrespective of the collision system in which $\Lambda$ hyperons are produced, they are also expected to contribute to the observed $\Lambda$ polarization in HICs. An important question then naturally arises: could these two phenomena in fact be related?

At first glance, these two measurements appear unrelated. Transverse $\Lambda$ polarization is defined with respect to the production plane, which is determined individually for each $\Lambda$ hyperon. In contrast, global $\Lambda$ polarization is measured relative to the reaction plane of the entire event and is driven by the system’s orbital angular momentum, independent of where a given $\Lambda$ hyperon is produced. However, the sizable anisotropic flow observed in HICs~\cite{PhysRevLett.120.062301,PhysRevC.103.034908,2022137003}, particularly the directed flow, strongly correlates $\Lambda$ emission with the reaction plane. As a consequence, this pronounced momentum-space anisotropy can induce a nontrivial correlation between the $\Lambda$ production plane and the reaction plane.

\begin{figure*}[t]
    \centering
    \includegraphics[width=0.85\textwidth]{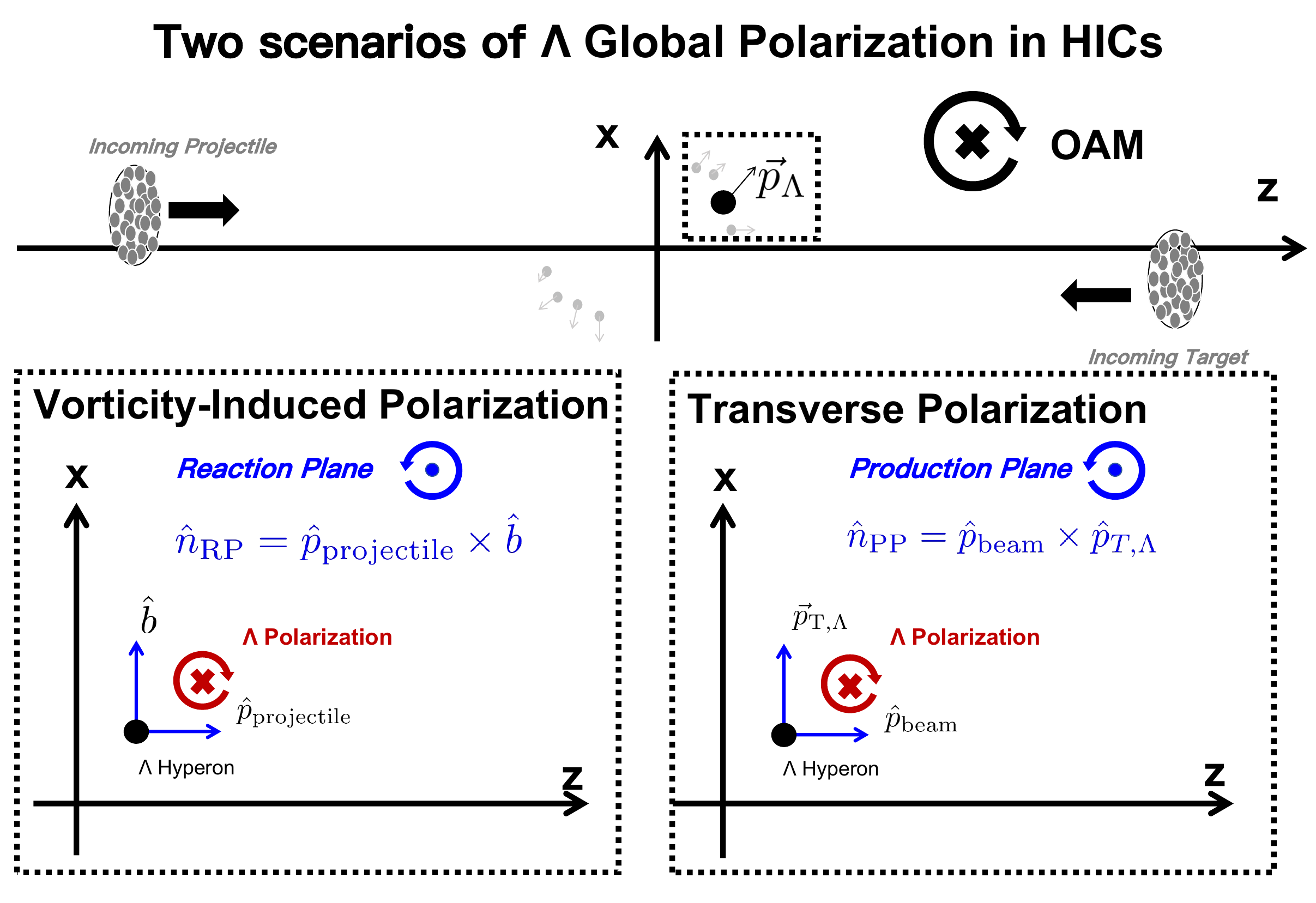}
    \caption{Illustration of heavy ion collisions (top) and two scenarios of $\Lambda$ global polarization: vorticity-induced $\Lambda$ global polarization (bottom left) and that induced by transverse polarization (bottom right). The black cross denotes the orbital angular momentum pointing into the page. Red crosses denote $\Lambda$ spins pointing into the page, while blue dots denote the normal vectors of the reaction and production planes pointing out 
of the page. }
    \label{fig:figure1}
\end{figure*}

In this paper, we explore this idea by proposing a distinct physics mechanism where $\Lambda$ global polarization emerges from the coupling of the transverse $\Lambda$ polarization and anisotropic flows. Specifically, i) we simulate low-energy HICs using the Jet AA Microscopic Transportation Model 2 (JAM2) \cite{PhysRevC.106.044902,PhysRevC.61.024901} event generator that has no vorticity-induced global $\Lambda$ polarization; ii) we introduce transverse $\Lambda$  polarization based on the world data to the JAM2 model particle by particle given their production planes and corresponding kinematics; iii) we re-analyze the modified simulation to measure the $\Lambda$ global polarization in the same way as the experimental data were analyzed; and finally iv) compare the results to the published data.

This paper is organized as follows. In Sec.~\ref{sec:pol}, we review the two distinct $\Lambda$ polarization mechanisms, and explicitly layout how they can be related via anisotripic flow in HICs. In Sec.~\ref{sec:model}, we introduce the Monte Carlo simulation and how our proposed mechanism gets implemented, followed by the quantitative results and their discussions in Sec.~\ref{sec:results}. Conclusions will be given in Sec.~\ref{sec:conclusion}.  

\section{Mechanism Description\label{sec:pol}}

Due to the self-analyzing weak decay of $\Lambda \rightarrow p + \pi^{-}$, the $\Lambda$  spin orientation can be inferred from the decay daughters' angular distribution. That is, in the $\Lambda$ hyperon's rest frame, the daughter proton exhibits the following angular distribution: 
\begin{equation}
    \frac{d N}{ d\Omega} = \frac{1}{4\pi} \left(1 +\alpha_{\Lambda} P_{\Lambda}~\hat{n} \cdot  \hat{p}\right),
    \label{eq:lambda_polarization}
\end{equation}
where $\alpha_{\Lambda} = 0.747 \pm0.009$  \cite{PhysRevD.110.030001} is the weak decay constant, $\hat{n}$ is a unit vector describing the direction of $\Lambda$ polarization, $P_{\Lambda}$ captures the magnitude of the polarization with respect to direction~$\hat{n}$, $\hat{p}$ is the unit vector of daughter proton's momentum in $\Lambda$ hyperon's rest frame, and the $\Omega$ is the angle between $\hat{n}$ and $\hat{p}$. 
\subsection{Transverse $\Lambda$ Polarization}
In the measurement of transverse $\Lambda$  polarization, the unit vector $\hat{n}$ is chosen as the normal direction of the production plane (PP), i.e., 
\begin{equation}
    \hat{n}_{\text{PP}} = \hat{p}_{\text{beam}} \times \hat{p}_{\Lambda},
    \label{eq:n_pp}
\end{equation}
where $\hat{p}_{\text{beam}}$ is the direction of the beam momentum, and $\hat{p_{\Lambda}}$ is the direction of $\Lambda$ hyperon's momentum in the laboratory frame. In asymmetric collision systems, beam direction $\hat{p}_{\text{beam}} $ can be defined unambiguously. In symmetric collision systems, we adopt the convention used in Ref.~\cite{PhysRevD.91.032004} that beam direction $\hat{p}_{\text{beam}}$ always points in the same direction as the longitudinal momentum of the associated $\Lambda$ hyperon. With this convention, the transverse $\Lambda$ polarization is observed to be negative relative to $\hat{n}_{\text{PP}}$.

\subsection{$\Lambda$ Global Polarization}
In heavy ion collisions, the $\Lambda$ global polarization is measured with respect to the reaction plane (RP) of the collision system. In parallel to the case of transverse $\Lambda$  polarization, the polarization axis of $\Lambda$ polarization can be written as~\cite{PhysRevLett.94.102301}: 
\begin{equation}
    \hat{n}_{\text{RP}} =\hat{p}_{\text{projectile}} \times \hat{b}
\end{equation}
where $\hat{b}$ is the direction of impact parameter  pointing from the center of target nucleus to the center of the projectile nucleus,  and $\hat{p}_{\text{projectile}}$ is the direction of projectile nucleus' momentum. It should be noted that the direction $\hat{n}_{\text{RP}}$ is $opposite$~\footnote{Note that although one can define the $\hat{n}_{\text{RP}}$ aligns with the OAM of the system, this convention was chosen specifically in Liang and Wang's original proposal~\cite{PhysRevLett.94.102301} to mimic the same negative polarization to the reaction (production) plane.} to the direction of system's orbital angular momentum, and therefore the $\Lambda$ global polarization should be also negative with respect to $\hat{n}_{\text{RP}}$. 

To clearly indicate the alignment between $\Lambda$ global polarization and system's orbital angular momentum, an additional minus sign is put in the definition of the observable. For practical reasons, the experimentally accessible event plane (EP) is used as a proxy for the reaction plane, and measured signals are corrected for the finite event-plane resolution. Therefore, the $\Lambda$ global polarization $\bar{P}_{\Lambda}^{(\text{GP})}$ can be written as~\cite{PhysRevC.101.044611,Abelev_2007},
\begin{equation}
    \bar{P}_{\Lambda}^{(\text{GP})} = -\frac{8}{\pi\alpha_{\Lambda}}\frac{1}{R_{\text{EP}}}\langle \sin\left(\phi_p^{*} -\Psi_{\text{EP}} \right) \rangle,
    \label{eq:GlobalPolarization}
\end{equation}
where $\phi_p^*$ is the azimuthal angle of daughter proton in the $\Lambda$ hyperon's rest frame, $\Psi_{\text{EP}}$ is the azimuthal angle of event plane, and $R_{\text{EP}}$ is the resolution of event plane. The angle brackets denote an average over all selected $\Lambda$ hyperons of all selected events. 


\subsection{Directed flow couples $\Lambda$ production to reaction plane}

Similar definitions of polarization axes $\hat{n}_{\text{PP}}$ and $\hat{n}_{\text{RP}} $ suggest that the correlation between the two can be developed, if there is a correlation between $\Lambda$ hyperon's transverse momentum and the impact parameter of the event, i.e., the reaction plane. Such correlation can be introduced by the anisotropic flow in HICs, where the particle azimuthal angle $\phi$ follows such a distribution\cite{flow_paper},
\begin{equation}
    \frac{dN}{d\phi} \propto 1 + 2 \sum_{n=1}^{\infty} v_n \cos n (\phi- \Psi_\text{{RP}}),
    \label{eq:flow}
\end{equation}
where $v_n$ characterizes the direction and magnitude of anisotropic flow. In particular, the directed flow $v_1$ can transfer the transverse $\Lambda$ polarization with respect to the production plane to the global polarization.

In Fig.~\ref{fig:figure1}, this mechanism is qualitatively illustrated. The coordinate frame is defined such that the $z$ axis follows the beam direction, with the positive (negative) rapidity side corresponding to the projectile (target) nucleus. The $x$ axis is chosen along the impact parameter vector $\hat{b}$ pointing from the target to the projectile nucleus, so that the reaction-plane azimuth is $\Psi_{\rm RP}=0$, while $\hat{y}=\hat{z}\times\hat{x}$. In this frame, the orbital angular momentum of the system points in the $-\hat{y}$ direction. In low-energy HICs, the directed flow $v_1$ in Eq.~\ref{eq:flow} is positive (negative) for $\Lambda$ hyperons at positive (negative) rapidity, indicating that the transverse momentum preferentially points toward $+(-)\hat{x}$ for positive (negative) rapidity. With the convention $\hat{p}_{\rm beam}=+(-)\hat{z}$ for positive- (negative-) rapidity $\Lambda$ hyperons, Eq.~\ref{eq:n_pp} implies that the transverse polarization axis $\hat{n}_{\rm PP}$ always points along $+\hat{y}$. Since the transverse polarization is negative with respect to $\hat{n}_{\rm PP}$, the $\Lambda$ spin is preferentially oriented along $-\hat{y}$, coinciding with the direction of the system's orbital angular momentum. 

This demonstrates that the transverse $\Lambda$  polarization, through its coupling to $v_1$ flow, can manifest itself in the measurement of the global polarization. Note that according to Eq.~\ref{eq:n_pp}, the beam direction $\hat{p}_{\text{beam}}$ is defined along the longitudinal momentum of the $\Lambda$, introducing a rapidity-sign dependence in the transverse polarization axis $\hat{n}_{\text{PP}}$. Since the directed flow $v_1$ is odd in $\Lambda$ hyperon's rapidity $y_{\Lambda}$, it drives opposite preferential transverse momenta in positive and negative rapidities (see Eq.~\ref{eq:flow}), thereby compensating the $y_{\Lambda}$-sign dependence introduced by the definition of $\hat{p}_{\text{beam}}$. As a result, only a single alignment between $\hat{n}_{\text{PP}}$ and $\hat{n}_{\text{RP}}$ remains, independent of $y_{\Lambda}$. In contrast, the elliptic flow $v_2$, being even in rapidity, generates two opposite alignment configurations with equal probability and thus cannot transfer transverse polarization into a net global polarization.

That said, measurements in high-energy HICs indicate that the directed flow $v_1$ reverses sign and has a much smaller magnitude than that in low-energy HICs. Therefore, the mechanism described above cannot be applied to HICs above 14.5 GeV~\cite{PhysRevLett.120.062301}. 

\begin{figure}[t]
    \centering
    \includegraphics[width=0.5\textwidth]{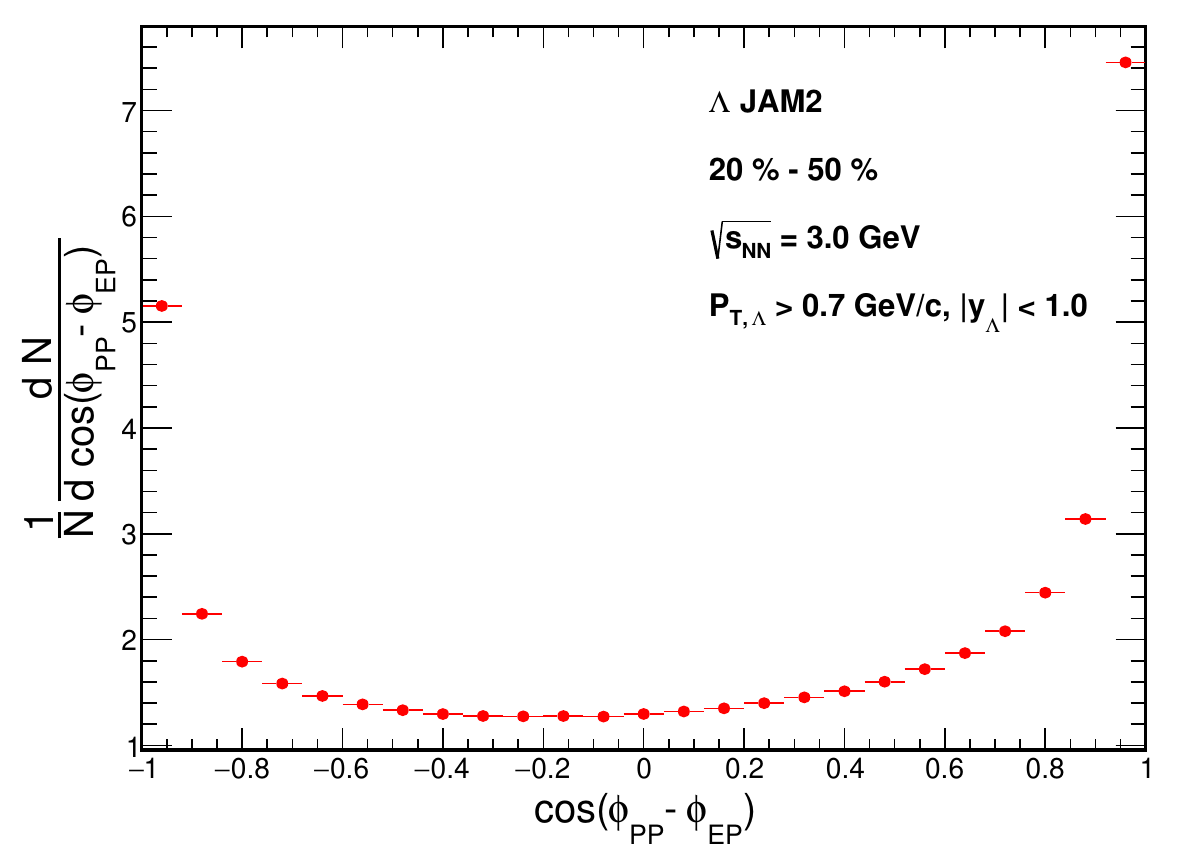}
    \caption{Distribution of $\cos (\phi_{\text{PP} } - \phi_{\text{EP}} )$ for Au-Au collisions at $\sqrt{s_{\rm{NN}}}= 3~\text{GeV}$ simulated by JAM2 generator.  \label{fig:cos_phi_pp_psi}}
\end{figure}

\begin{figure}[t]
    \centering
    \includegraphics[width=0.5\textwidth]{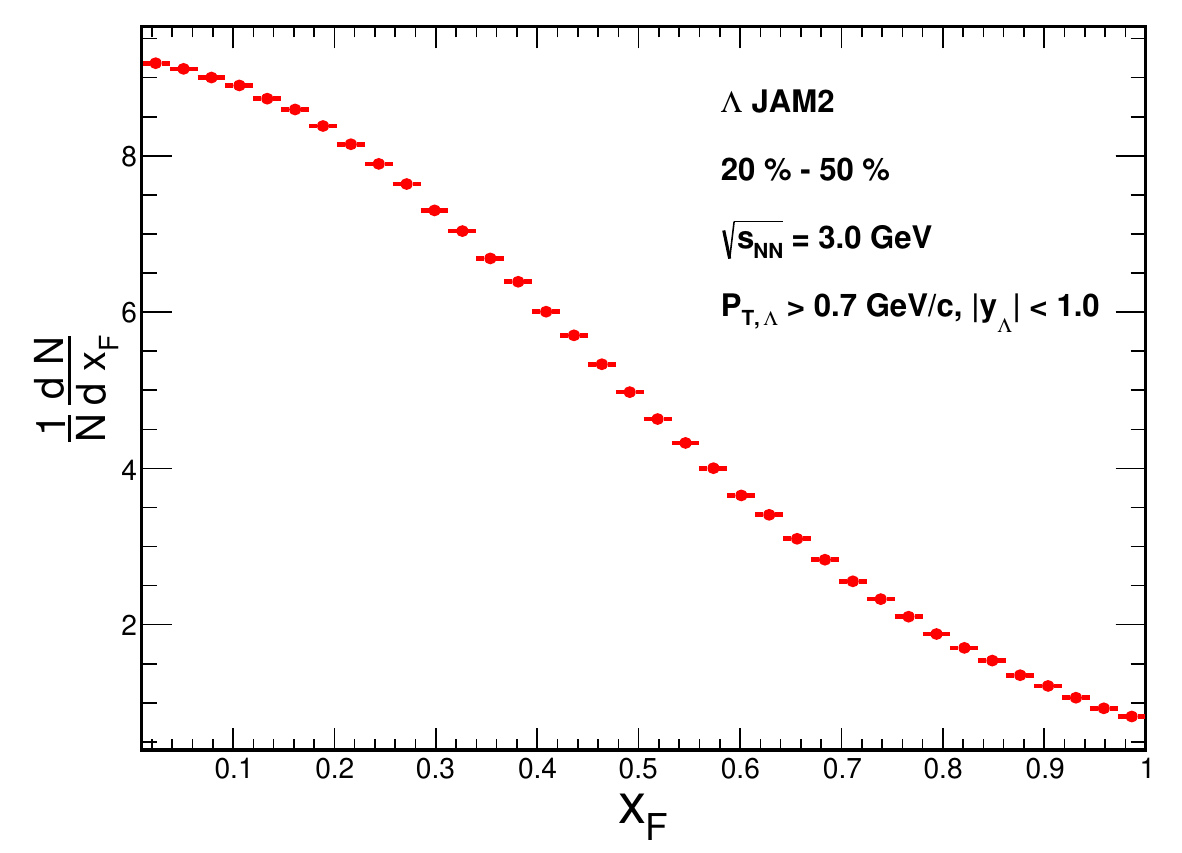}
    \caption{The distribution of Feynman-$x$($x_F$) of $\Lambda$ hyperons produced in Au-Au collisions at $\sqrt{s_{\rm{NN}}} = 3~\text{GeV} $. The result is obtained from events generated by JAM2.}
    \label{fig:xF_distribution}
\end{figure}


\section{Quantitative Modeling\label{sec:model}}


To quantify the above effect, we study the gold-gold (Au$+$Au) collisions at $\sqrt{s_{_{\rm{NN}}}}=3~\text{GeV}$ to estimate the resulted magnitude of $\Lambda$ global polarization. First, we employ the JAM2 event generator to simulate heavy-ion collisions. $\Lambda$ potential derived from the chiral effective field theory is implemented to the JAM2 event generator \cite{PhysRevC.106.044902,PhysRevC.61.024901} such that  $v_1$ flow of $\Lambda$ hyperons can be realistically reproduced as observed by the STAR collaboration~\cite{2022137003}. 

The transverse $\Lambda$  polarization is not incorporated in the JAM2 event generator, but can be implemented via a kinematic-dependent reweighting technique. The procedure is as follows: 

\begin{enumerate}
    \item Production plane determination. The normal direction of the production plane is calculated using Eq.~\ref{eq:n_pp}. The resulted $\hat{n}_{\text{PP}}$ determines the direction of $\Lambda$ polarization in its rest frame. 
    \item Transverse polarization determination. Previous experiments have established that transverse $\Lambda$  polarization exhibits a strong dependence on both transverse momentum ($p_{\Lambda,\text{T}}$) and Feynman-$x$ ($x_F$) of $\Lambda$ hyperons. In this study, however, we treat the polarization $P^{\text{(TP)}}_{\Lambda}$ as a function of $x_F$ obtained from global data fit~\cite{PhysRevC.109.055205}, neglecting its $p_\text{T}$ dependence for simplicity. 
    \item Weight factor determination. Since $\Lambda$ polarization is encoded in its decay angular distribution (see Eq.~\ref{eq:lambda_polarization}), $\Lambda$ hyperon is first decayed uniformly in its rest frame, and weighted by a factor of $\frac{1}{2} \left( 1 + \alpha_{\Lambda} P_{\Lambda}^{(\text{TP})} \cos \theta^*  \right)$ in the subsequent calculation of observables. Here, $\theta^*$ is the relative angle between normal vector of production plane $\hat{n}_{\text{PP}}$ and daughter proton's momentum vector in $\Lambda$ hyperon's rest frame. 
\end{enumerate}
After applying this reweighting technique, the transverse $\Lambda$  polarization becomes naturally coupled to its $v_1$ flow. 

For a direct comparison with STAR data, the same kinematic selections as those in the measurement \cite{PhysRevC.104.L061901} are used. Rapidity $y$ is defined in the collision center-of-momentum frame, while pseudorapidity $\eta$ is defined in the fixed-target laboratory frame. The centrality is characterized by the number of charged particles with $-2  < \eta <   0$ and only events with  $20-50\%$ centrality enter the following analysis. The first-order event plane $\Psi^{(1)}_{\text{EP}}$ is estimated by using charged particle with $-2.84  <\eta<  -2.55$, and the corresponding resolution $R_{\text{EP}}^{(1)}$ is determined using three-subevent method, where the two reference subevents are calculated using charged particles with $-0.5  <\eta<  -0.4$ and $-0.2  <\eta<  -0.1$. $\Lambda$ hyperons generated by JAM2 generator are selected with $p_{\Lambda,\text{T}} > ~0.7 ~\text{GeV}/c $ and $-0.2  <y_{\Lambda}< 1.0$ in the calculation of global polarization observable $\bar{P}_{\Lambda}^{(\text{GP})}$.

Correlation between the $\Lambda$ hyperon's production plane and the event plane can be captured by a quantity, namely, $\cos\left(  \phi_{\text{{{PP}}}} - \phi_{\text{EP}} \right)$, where $\phi_{\text{PP}}$ and $\phi_{\text{EP}}$ are the azimuthal angles of normal vectors of production plane and event plane, respectively. The distribution of $\cos\left( \phi_{\text{PP}} - \phi_{\text{EP}}\right)$ is shown in Fig.~\ref{fig:cos_phi_pp_psi}, where $\Lambda$ hyperons are selected with rapidity $|y_{\Lambda}|<1.0$.  This distribution exhibits a significant asymmetry, with a higher yield on the positive side than the negative side. This asymmetry implies an alignment configuration between the production plane and reaction plane (event plane), i.e., aligned normal vectors configuration $\hat{n}_{\text{PP}}\cdot\hat{n}_{\text{RP}} > 0 $. 

\begin{figure}[t]
    \centering
    \includegraphics[width=0.5\textwidth]{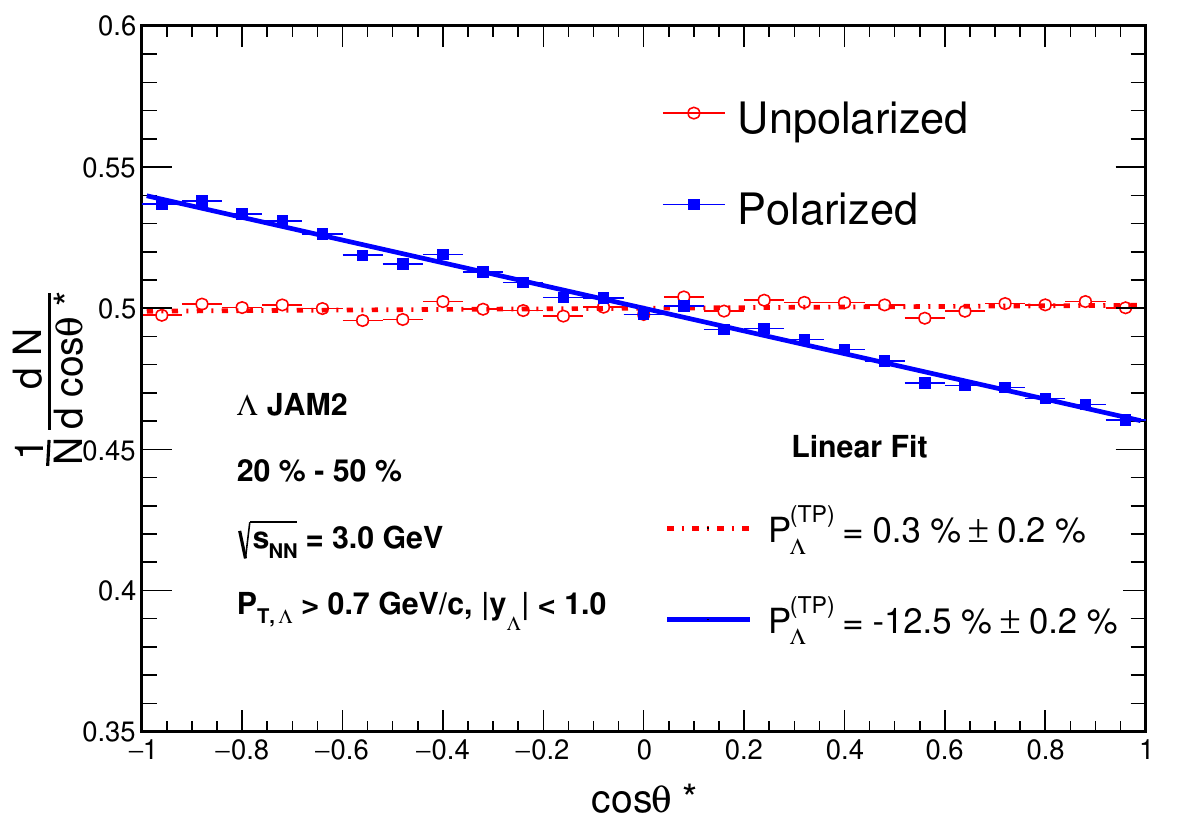}
    \caption{Normalized distribution of $\cos\theta^*$, where $\theta^*$ is the angle between the normal vector of the $\Lambda$ production plane $\hat{n}_{\text{PP}}$ and the momentum of the daughter proton. Red circles show the distribution without polarization, while blue squares correspond to the case with implemented transverse polarization. The lines represent linear fits to the corresponding distributions: red dashed line for the unpolarized case and blue solid line for the polarized case. The results are obtained from events generated by JAM2.}
    \label{fig:transverse_polarization}
\end{figure}

Figure.~\ref{fig:xF_distribution} shows the $x_F$ distribution of $\Lambda$ hyperons with $|y_{\Lambda}| < 1.0$ at $\sqrt{s_{_{\rm{NN}}}} = 3~ \text{GeV}$. With the convention that $\hat{p}_{\text{beam}}$ takes the same direction as longitudinal momentum of  $\Lambda$ hyperon, $x_F$ is calculated as positive for both $\Lambda$ hyperon with $y_{\Lambda} > 0$ and  $y_{\Lambda} < 0$. Subsequently, the magnitude of the transverse $\Lambda$  polarization is implemented as a function of $x_F$, i.e., $P_{\Lambda}^{(\text{TP})} (x_F)$. 
The function $P_{\Lambda}^{\text{(TP)}}(x_F)$ is obtained from a quadratic fit to the global data~\cite{PhysRevD.91.032004,ABT2006415,PhysRevD.40.3557}, which is the same as that used in~Ref.~\cite{PhysRevC.109.055205}. The low collision energy allows $\Lambda$ hyperons to populate the large-$x_F$ region, where the magnitude of transverse polarization is known to be substantial based on global data. 

Consequently, a significant transverse polarization can be expected in our simulation, as presented in Fig. \ref{fig:transverse_polarization}. Here, the angle $\theta^*$ is the relative angle between normal vector  of $\Lambda$ hyperon's production plane $\hat{n}_{\text{PP}}$ and direction of daughter proton's momentum in the rest frame of $\Lambda$ hyperon. The red circle markers depict the $\Lambda$ isotropic decay without any build-in polarization, which can serve as a  baseline scenario. The blue square markers, on the other hand, are derived by applying the reweighting technique to incorporate transverse polarization. The magnitude of $x_F$-integrated transverse polarization can be extracted from a linear fit. With the $x_F$ allowed by the selected kinematic ranges, the magnitude of $P_{\Lambda} ^{(\text{TP})}$ is found to be $-12.5 \% \pm 0.2 \% $. 

\section{Results and Discussion\label{sec:results}}
\begin{figure*}[th]
    \centering
    \includegraphics[width=0.9\textwidth]{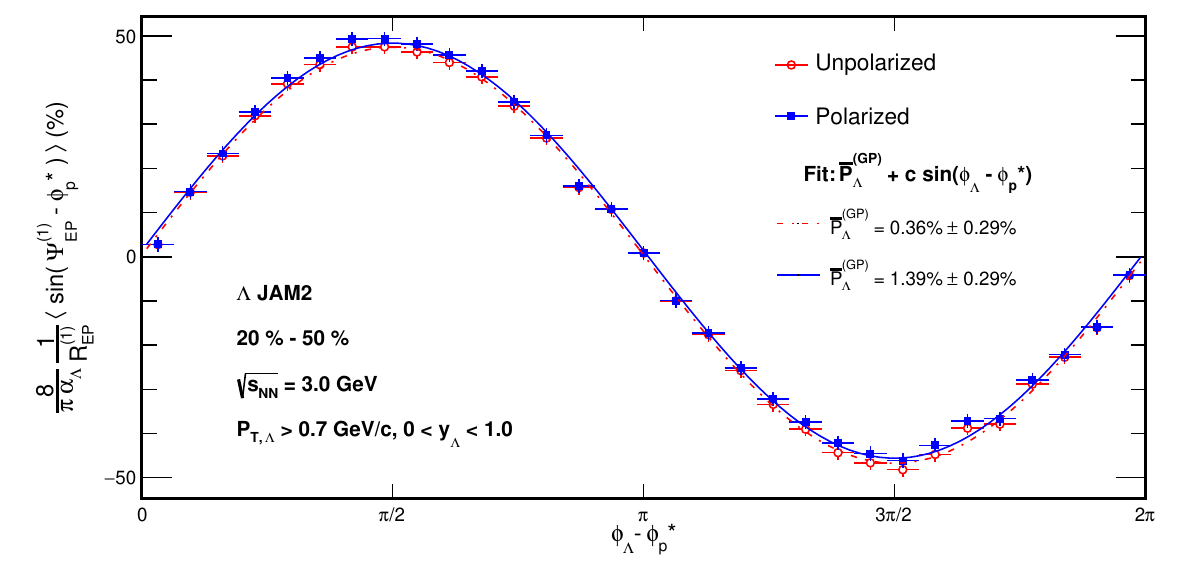}
    \caption{The global polarization calculated as a function of $\phi_{\Lambda}-\phi_{p}^*$, where $\phi_{\Lambda}$ is the azimuthal angle of $\Lambda$ hyperon in laboratory frame, and $\phi_p^*$ is the azimuthal angle of daughter proton in $\Lambda$ hyperon's rest frame. Red circles show the distribution without polarization, while blue squares correspond to the case with implemented transverse polarization. The lines represent fits to the corresponding distributions: red dashed line for the unpolarized case and blue solid line for the polarized case. The results are obtained from events generated by JAM2.}
    \label{fig:STAR_method}
\end{figure*}

The simulation study based on the JAM2 model has produced around 80 million minimum-bias events. We first calculate the $\Lambda$ global polarization using Eq. \ref{eq:GlobalPolarization}, which is a direct average over all selected $\Lambda$ hyperons from all selected events. It is found that the global polarization is $23\% \pm 6\%$.


However, a more sophisticated method has been utilized to extract the $\Lambda$ global polarization in low-energy HICs~\cite{PhysRevC.104.L061901} in the STAR experiment. Specifically, the global polarization is first calculated as a function of $\phi_\Lambda-\phi_p^*$, where $\phi_{\Lambda}$ is the azimuthal angle of $\Lambda$ hyperon in the lab frame, and $\phi_{p}^{*}$ is the azimuthal angle of daughter proton in rest frame of the $\Lambda$ hyperon. In each bin of $\phi_\Lambda-\phi_{p}^*$, the $\Lambda$ global polarization is calculated using Eq.~\ref{eq:GlobalPolarization}, and then a fit to the following function, i.e.,
\begin{equation}
    \frac{8}{\pi \alpha_\Lambda} \frac{1}{R_{\text{EP}}^{(1)}} \langle \sin(\Psi_{\text{EP}}^{(1)}-\phi_p^*)\rangle= \bar{P}_{\Lambda}^{(\text{GP})} + c~\sin(\phi_\Lambda-\phi_p^* ),
    \label{eq:fit_GP}
\end{equation}
is performed. The vertical shift $\bar{P}_{\Lambda}^{\text{(GP)}}$ corresponds to the signal of  $\Lambda$ global polarization. The presence of a sinusoidal function originates from the strong $v_1$ flow, with the amplitude $c$ proportional to rapidity-integrated $v_1$ flow. For detailed experimental procedure, see Ref.~\cite{PhysRevC.104.L061901}. 
This analysis procedure is also performed in this study, compared with the one calculated as a direct average. By using this method, Fig. \ref{fig:STAR_method} shows the polarization extracted for $\Lambda$ hyperons with rapidity $0<y_{\Lambda}<1.0$. The red circle markers represent $\Lambda$ hyperons undergoing isotropic decay, while the blue square markers correspond to $\Lambda$ hyperons with implemented transverse polarization. The presence of transverse polarization modifies the sinusoidal shape, enhancing the maximum at $\pi/2$ and flatting the minimum at $3\pi/2$. This can be explained by that transverse polarization of the $\Lambda$ hyperon causes $\phi_{\Lambda}$, the azimuthal angle of $\Lambda$ hyperon, to preferentially align at an angle of $\pi/2$ relative to $\phi_p^*$ , the azimuthal angle of daughter proton in $\Lambda$ rest frame. As a result, an enhanced vertical shift can be produced from this distortion. 

\begin{figure*}[ht]
    \centering
    \includegraphics[width=\textwidth]{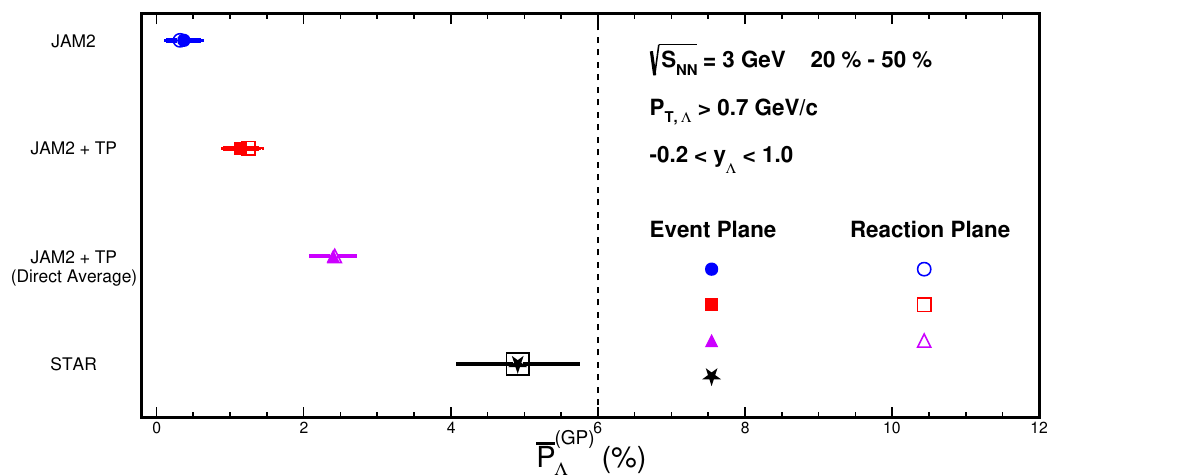}
    \caption{The $\Lambda$ global polarization $P_{\Lambda}^{(\text{GP})}$ extracted from JAM2 events and STAR data for Au+Au collisions at $\sqrt{s_{\rm{NN}}} = 3~\text{GeV}$. The blue circle markers denote $\Lambda$ hyperons without any polarization. Red square markers indicate the case with implemented transverse polarization of $\Lambda$ hyperons, and the $P_{\Lambda}^{(\text{GP})}$ is extracted as an offset in Eq.~\ref{eq:fit_GP}. The violet triangle markers also indicate the case of polarized $\Lambda$ hyperons, but the global polarization is extracted as a direct average using Eq.~\ref{eq:GlobalPolarization}. The black star maker shows the $P_{\Lambda}^{(\text{GP})}$ measured by STAR collaboration. The $P_{\Lambda}^{(\text{GP})}$ calculated respect to event plane is displayed by solid marker, while the one calculated with respect to reaction plane is shown in hollow marker. The systematic uncertainty of STAR measurement is represented with box.}
         \label{fig:TP_GP}
\end{figure*}

The analysis procedure is performed for $\Lambda$ hyperons with $0<y_{\Lambda}<1.0$ and $-0.2<y_{\Lambda}<0$, separately, and eventually a weighted average is extracted, as shown in Fig.~\ref{fig:TP_GP}. The global polarization extracted by fitting to Eq.~\ref{eq:fit_GP} of  $\Lambda$ hyperons undergoing isotropic decay is displayed in blue circle markers, and it is basically consistent with zero within the statistical uncertainty.  With implemented transverse polarization (abbreviated as ``TP''), $\Lambda$ hyperons exhibit a large global polarization accounting for about $49\% \pm 10\%$ of STAR measurement (shown in black star marker) if extracted by taking direct average as Eq.~\ref{eq:GlobalPolarization} (shown in violet triangle marker). However, if extracted by fitting to Eq.~\ref{eq:fit_GP} (shown in red square marker), the global polarization is reduced to approximately $23\%\pm 6\% $ of the value reported by the STAR collaboration. In addition, the global polarization calculated relative to the reaction plane, shown with hollow markers, agrees well with the event-plane results. 

Under the same kinematic selections and analysis procedures as those employed in the experimental measurements, the mechanism proposed in this study is capable of generating a sizable $\Lambda$ global polarization even in the absence of QGP vorticity. This implies that current measurements, which cannot disentangle polarization induced by vorticity from that arising from transverse polarization, may contain a non-negligible contribution from the latter. Therefore, in low-energy HICs, the observed $\Lambda$ global polarization, commonly interpreted as a probe of QGP vorticity, should be treated with caution, particularly when compared with theoretical calculations~\cite{PhysRevC.104.L041902,PhysRevC.103.L031903}.

That said, in high-energy HICs, the magnitude of $v_1$ flow of $\Lambda$ hyperons is significantly reduced, leading to much weaker alignment between production plane and reaction plane. It can be expected that contributions from transverse polarization become negligible. 

Finally, here are a few remarks on this study:
\begin{itemize}
    \item Detector simulation was not carried out, because it requires access to collaboration software/resources. Although it is not expected to be large, experiments are encouraged to estimate this effect based on their detector conditions. 
    \item In the original proposal of $\Lambda$ global polarization~\cite{PhysRevLett.94.102301}, it was predicted that transverse $\Lambda$ polarization may enhance the observed $\Lambda$ global polarization in the rapidity region where the $v_1$ flow is strong. The rapidity dependence of the proposed effect is the same as that in vorticity-driven polarization~\cite{Liang:2019pst} - it grows as rapidity increases because $P_{\Lambda}^{\text{(TP)}}(x_F)$ grows with $x_F$ (or $y$). That said, a quantitative study can be performed in the future and the rapidity dependence could be compared. 
    \item One of the strongest evidences to set the transverse $\Lambda$ polarization apart from vorticity-driven global polarization is the nonzero $\bar{\Lambda}$ polarization in HICs. Although the transverse $\bar{\Lambda}$ polarization has always been found consistent with zero, the statistical uncertainty in low-energy experiments is at a few percent level~\cite{SELEX:2007bjj,PhysRevD.40.3557,WA89:1994nzv,ABT2006415,PhysRevD.76.092008,ASTIER20013}, which can be comparable to the magnitude observed from HICs. If the transverse polarization of $\bar{\Lambda}$ has the opposite sign to that of $\Lambda$, the proposed mechanism naturally yields similar global polarization for $\bar{\Lambda}$ and $\Lambda$. Otherwise, a significant difference is expected at $\sqrt{s_{\mathrm{NN}}}= 3~\text{GeV}$ but becomes negligible compared to precision of current measurements \cite{Hu:2024suz} at higher collision energies due to the much weaker directed flow.

    On the flip side, if this mechanism can be further quantified in the experiment, this might be the first-time we have observed an nonzero transverse $\bar{\Lambda}$ polarization, which would solve part of the 50-year puzzle.

    \item Another related and intriguing phenomenon is the local $\Lambda$ polarization in the beam ($z$) direction in HICs~\cite{STAR:2019erd,ALICE:2021pzu} and recently in proton-nucleus collisions~\cite{CMS:2025nqr}. It is obvious that our mechanism cannot simply applied. However, recently STAR Collaboration has reported the first transverse $\Lambda$ polarization in jet with a modified production plane - $\hat{n'} = \hat{p}_{\text{jet}} \times \hat{p}_{\Lambda}$ to study the effect from fragmentation~\cite{STAR:2025jwc}. Whether these two effects are related would be an interesting follow-up to our proposed mechanism. 
    
\end{itemize}

\section{Summary\label{sec:conclusion}}
In conclusion, we have proposed a novel mechanism for $\Lambda$ global polarization in heavy-ion collisions. This mechanism is based on the idea that the transverse polarization of $\Lambda$ hyperons, long observed in various collision systems, can contribute to the measured global polarization signal. In the presence of large directed flow at low beam energies, the $\Lambda$ production plane becomes correlated with the reaction plane, leading to an apparent global polarization aligned with the system’s orbital angular momentum. To quantify this effect, we have developed a dedicated modeling framework and performed realistic simulations of Au+Au collisions using the JAM2 Monte Carlo event generator at $\sqrt{s_{\rm NN}} = 3$~GeV. Our results show that this mechanism can account for approximately $23\%\pm6\%$ of the global polarization signal reported by the STAR Collaboration. Since current analysis techniques cannot disentangle the contribution arising from transverse $\Lambda$  polarization from that induced by QGP vorticity, the former represents a non-negligible effect to vorticity-driven polarization measurements. The present study provides a quantitative estimate of this mechanism, which is essential for a reliable interpretation of experimental data.

\begin{acknowledgments}
The authors would like to thank Ashik Ikbal and Sooraj Radhakrishnan for discussions of JAM2 event genarator, and Arjun Kumar, Shusu Shi, Shuzhe Shi, Prithwish Tribedy, Xin-Nian Wang, Li Xu, Nu Xu, Zhangbu Xu for discussions of the paper draft. We also like to thank the Center for Frontier Nuclear Science (CFNS) at Stony Brook University and the Electron-Ion Collider (EIC) group at Brookhaven for informal discussions. The work of F.~Liu and Z.~Tu are supported by the U.S. Department of Energy under Award DE-SC0012704 and the BNL Laboratory Directed Research and Development (LDRD) 26-029 project.
\end{acknowledgments}

\section*{Data Availability Statement}
The results presented in this work were obtained using the publicly available JAM2 event generator \cite{PhysRevC.106.044902,PhysRevC.61.024901}.

\bibliographystyle{apsrev4-2}
\bibliography{reference}
\end{document}